\documentclass[twoside,preprintnumbers,amsmath,amssymb,pacs,nofootinbib]{revtex4}
\usepackage{graphicx}
\usepackage{dcolumn}
\usepackage{bm}
\usepackage{float}
\usepackage{txfonts}
\usepackage{pslatex}
\usepackage{amssymb}
\usepackage[parfill]{parskip}
\usepackage[Gray,squaren,thinqspace,thinspace]{SIunits}
\usepackage{graphicx}
\usepackage[small,bf]{caption}
\DeclareMathOperator{\tr}{tr}
\newcommand{\lms}{\Lambda_{\overline{\text{MS}}}}

\begin{document}

\title{\textbf{The asymmetry of the dimension 2 gluon condensate: the finite temperature case}}
\author{David~Vercauteren$^a$, Henri~Verschelde$^a$} \email{David.Vercauteren@UGent.be, Henri.Verschelde@UGent.be}
 \affiliation{\vskip 0.1cm
                            $^a$ Ghent University, Department of Physics and Astronomy \\
                            Krijgslaan 281-S9, B-9000 Gent, Belgium\\\\\vskip 0.1cm
}

\begin{abstract}
In this paper, we continue the work begun in a previous article. We compute, in the formalism of local composite operators, the value of the asymmetry in the dimension two condensate for finite temperatures. We find a positive value for the asymmetry, which disappears when the temperature is increased. We also compute the value of the full dimension two condensate for higher temperatures, and we find that it decreases in abolute value, finally disappearing for sufficiently high temperature. We also comment on the temperature dependence of the electric and magnetic components of the condensate seperately. We compare our results with the corresponding lattice date found by Chernodub and Ilgenfritz.
\end{abstract}
\maketitle

\section{Introduction}
The dimension 2 gluon condensate $\langle A_\mu^2\rangle$ in pure Yang-Mills theory has been proposed in \cite{Gubarev:2000eu,Gubarev:2000nz}, and it has been investigated in different ways since then \cite{Verschelde:2001ia, Dudal:2002pq, Dudal:2003vv, Vercauteren:2007gx, Dudal:2005na,Boucaud:2001st, Furui:2005he, Gubarev:2005it, Browne:2003uv, Andreev:2006vy, RuizArriola:2006gq, Chernodub:2008kf}.

In \cite{Verschelde:2001ia} an analytical framework for studying this condensate has been developed, based on work carried out in the Gross-Neveu model \cite{Verschelde:1995jj}. Different problems had to be overcome. First of all there is the gauge invariance of this condensate. In order to make the operator $A_\mu^2$ gauge invariant, one can take the minimum of its integral over the gauge orbit. Since $\int d^dx\, A_\mu^U A_\mu^U$, with $U\in SU(N)$, is positive, this minimum will always exist. In a general gauge, however, the minimum is a highly nonlocal and thus hard to handle expression of the gauge field. A minimum is however reached in the Landau gauge ($\partial_\mu A_\mu=0$), though, so that working in this gauge reduces the operator to a local expression\footnote{We ignore the Gribov problem here, see also \cite{Dudal:2005na}.}. Secondly adding a source $J$, coupled to $A_\mu^2$, makes the theory nonrenormalizable at the quantum level. To solve this, a term quadratic in the source must be added, which in turn spoils the energy interpretation of the effective action. One way around this is to perform the Legendre inversion, but this is rather cumbersome, especially so with a general, spacetime dependent source. One can also use a Hubbard-Stratonovich transform, which introduces an auxiliary field (whose interpretation is just the condensate) and eliminates the term quadratic in the source. Details can be found in \cite{Verschelde:2001ia}. The result was that the Yang-Mills vacuum favors a finite value for the expectation value of $A_\mu^2$. The precise renormalization details of the procedure proposed in \cite{Verschelde:2001ia} were given in \cite{Dudal:2002pq}.

Recently, Chernodub and Ilgenfritz \cite{Chernodub:2008kf} have considered the asymmetry in the dimension two condensate. They performed lattice simulations, computing the expectation value of the electric-magnetic asymmetry in Landau gauge, which they defined as
\begin{equation}\label{ci}
\Delta_{A^2} = \langle g^2 A_0^2 \rangle - \frac1{d-1} \sum_{i=1}^{d-1} \langle g^2 A_i^2 \rangle \;.
\end{equation}
At zero temperature, this quantity must of course be zero due to Lorentz invariance\footnote{We shall deliberately use the term Lorentz invariance, though we shall be working in Euclidean space throughout this paper.}. Necessarily it cannot diverge as divergences at finite $T$ are the same as for $T=0$, hence this asymmetry is in principle finite and can be computed without renormalization, for all temperatures.

A first remark concerns the visibility of the (de)confinement phase transition in the value of the asymmetry\cite{Chernodub:2008kf}. At temperatures lower than the critical one, the asymmetry goes from zero at zero temperature to a positive value, which reaches a maximum at the critical temperature. At higher temperatures, the asymmetry decreases and becomes negative when $T>2.21\;T_c$. The two transition points ---the phase transition temperature and the symmetric point where the asymmetry goes through zero--- divide the temperature range in three regions. These seem to coincide with those associated with the condensed, liquid, and gaseous states of the magnetic monopoles, whose dynamics are closely related to confinement and deconfinement (see, for example, \cite{Chernodub:2006gu}). At yet higher temperatures, one would expect the perturbative behavior to kick in, which goes like
\begin{equation} \label{deltapertverw}
\Delta_{A^2}(T) = \frac{N^2-1}{12}g^2T^2
\end{equation}
at lowest order.\footnote{In \cite{Chernodub:2008kf} the opposite sign was erroneously found, which seemed to agree with the highest temperatures found in the lattice computations. Given the sign of \eqref{deltapertverw}, one would expect the qualitative behavior of asymmetry to make yet another turn at higher temperatures.} However, lattice artifacts prohibit lattice computations at sufficiently high temperatures to see this \cite{maxim}.

At low temperatures, from thermodynamical arguments one would expect an exponential fall-off with the lowest glueball mass in the exponent, $\Delta\sim e^{-m_{\text{gl}}T}$. Instead, the authors of \cite{Chernodub:2008kf} found an exponential with a mass $m$ significantly smaller than $m_\text{gl}$. So far, there is no explanation for this behavior.

In \cite{wij}, the authors and collaborators have extended the framework from \cite{Verschelde:2001ia} to include the asymmetry $\Delta_{A^2}$. In that article, the potential was computed for $T=0$, and no non-trivial value for the asymmetry was found --- as is necessary for Lorentz invariance. In this paper we extend the computations in order to include finite temperature effects, with the aim of shedding more light on the results of \cite{Chernodub:2008kf}. In section \ref{prel} we give a short review of what was found in \cite{wij}, which is then continued by a computation of the finite temperature effective action in section \ref{tr}. In section \ref{min} we find and discuss the minima of the potential, the values of the different condensates and their temperature dependence. Section \ref{conc} concludes the paper.

\section{Preliminaries} \label{prel}
In \cite{wij} the effective action in the presence of a dimension two condensate and of an asymmetry was computed. Since the starting point for the calculations when $T\not=0$ are identical, we shortly review the steps taken in \cite{wij}.

The starting point to compute the effective potential is the following Lagrangian density:
\begin{eqnarray}\label{sigmaactie}
\mathcal L(A_\mu, \sigma, \phi_{\mu\nu}) &=&  \mathcal L_{YM} +  \mathcal L_\text{gf} + \frac1{2\zeta} \frac{\sigma^2}{g^2} +  \frac1{2\zeta g} \sigma A^a_\mu A^a_\mu + \frac1{8\zeta} (A^a_\mu A^a_\mu)(A^b_\nu A^b_\nu) \nonumber\\
&& + \frac1{2\omega}\frac{\varphi_{\mu\nu}^2}{g^2} +  \frac1{2\omega g}Z_A \varphi_{\mu\nu} A_\mu A_\nu + \frac1{8\omega} (A^a_\mu A^a_\nu)(A^b_\mu A^b_\nu) \;,
\end{eqnarray}
where the following couplings have been introduced:
\begin{eqnarray}
\zeta &=& \frac{N^2-1}{g^2N} \left(\frac9{13} + \frac{161}{52} \frac{g^2N}{16\pi^2} \right) \;, \\
\omega &=& \frac{N^2-1}{g^2N} \left(\frac{1}{4} + \frac{73}{1044} \frac{g^2N}{16\pi^2} \right) \;,
\end{eqnarray}
to one-loop order. The vacuum expecation values of the $\sigma$ and $\varphi_{\mu\nu}$ fields are
\begin{eqnarray}
\langle\sigma\rangle &=& -\frac g2 \langle A_\mu^2 \rangle \;, \\
\langle\varphi_{\mu\nu}\rangle &=& -\frac g2 \left\langle A_\mu^aA_\nu^a - \frac{\delta_{\mu\nu}}{d}A^a_\kappa A_\kappa^a\right\rangle \; .
\end{eqnarray}
In order to simplify notations, we set
\begin{eqnarray}
m^2 = g\sigma' &=& \frac{13}9\frac N{N^2-1} g\sigma \;, \\
M_{\mu\nu} = g\varphi_{\mu\nu}' &=& 4\frac N{N^2-1} g\varphi_{\mu\nu} \;,
\end{eqnarray}
which denote an effective mass and an effective mass matrix. With these notations, the condensates as defined with the conventions of \cite{Chernodub:2008kf} are
\begin{eqnarray}
\langle g^2A_\mu^2\rangle &=& - \frac{18}{13} \frac{N^2-1}N m^2 \;, \\
\Delta_{A^2} = \langle g^2A_0^2\rangle - \frac13 \langle g^2A_i^2\rangle &=& - \frac12 \frac{N^2-1}N M_{00} \;,
\end{eqnarray}
where the Latin index denotes the space components.

With these givens it is possible to compute the effective action for a space-time independent $\sigma'$ and $M_{\mu\nu}$ using the background field formalism. We separate the two fields into a classical part and quantum fluctuations, after which the fluctuations can be integrated out. Expanding the resulting path integral over $A_\mu$ to one-loop order gives
\begin{equation}
V_\text{eff} (\sigma',M_{\mu\nu}) = \mathcal L[A_\mu=0,\sigma',M_{\mu\nu}] + \frac{N^2-1}2 \tr\ln \left( -\partial^2\delta_{\mu\nu} + \left(1-\frac1\xi\right)\partial_\mu\partial_\nu + \delta_{\mu\nu} g\sigma' + M_{\mu\nu}\right) \;,
\end{equation}
where the limit $\xi\rightarrow0$ is implied, as we work in the Landau gauge. As we are interested in the asymmetry, we parametrize the mass matrix as
\begin{equation}\label{param}
M_{\mu\nu} = A \begin{pmatrix} 1 &&& \\ &-\frac{1}{d-1}&& \\&&\ddots&\\ &&&-\frac{1}{d-1} \end{pmatrix} \;;
\end{equation}
i.e. we preserve rotational invariance in the spatial part. With this form, the $\tr\ln$ in the effective potential can be split into different parts, and in the limit $\xi\rightarrow0$ we get
\begin{eqnarray} \label{tracespols}
V_\text{eff} (\sigma',M_{\mu\nu}) &=& \mathcal L[A_\mu=0,\sigma',M_{\mu\nu}] + \frac{N^2-1}2\tr\ln (-\partial^2) + \frac{N^2-1}2(d-2) \tr\ln\left(-\partial^2+m^2-\frac {A}{d-1}\right) \nonumber \\ && + \frac{N^2-1}2\tr\ln\left(-\partial^2+m^2 +A\left(1-\frac{d}{d-1}\frac{\partial_0^2}{\partial^2}\right)\right) \;.
\end{eqnarray}

\section{The traces} \label{tr}
At finite $T$ and in Euclidean space-time, the spectrum of $-\partial^2$ is discrete --- the eigenvalues are $4\pi^2T^2n^2+\vec k^2$ where $n\in\mathbb Z$ are the Matsubara frequencies and $\vec k$ is the momentum in the spacelike directions. It happens to be convenient to take the second and the last terms of \eqref{tracespols} together (mark that $\tr\ln-\partial^2$ in dimensional regularization does not vanish for finite $T$, but it gives a constant contribution to the energy), so that we have to compute the following traces:
\begin{subequations} \label{beginnetje} \begin{gather}
\frac{N^2-1}2(d-2) T\int \frac{d^{d-1}k}{(2\pi)^{d-1}} \sum_{n=-\infty}^{+\infty} \ln\left(4\pi^2T^2n^2+\vec k^2+m^2-\frac A{d-1}\right) \;, \\
\frac{N^2-1}2 T\int \frac{d^{d-1}k}{(2\pi)^{d-1}} \sum_{n=-\infty}^{+\infty}\ln\left(16\pi^4T^4n^4+4\pi^2T^2n^2\left(m^2+2\vec k^2-\frac A{d-1}\right)+\vec k^2(\vec k^2+m^2+A)\right) \;.
\end{gather} \end{subequations}
The sums can be computed using standard techniques. In order to have convergent sums, one first writes (for concreteness, consider the first sum):
\begin{equation}
\sum_{n=-\infty}^{+\infty} \ln\left(4\pi^2T^2n^2+\vec k^2+m^2-\frac A{d-1}\right) = \left. \int d\mu \sum_{n=-\infty}^{+\infty} \frac1{4\pi^2T^2n^2+\vec k^2+m^2-\frac A{d-1}+\mu} \right|_{\mu=0} \;.
\end{equation}
Then the sum can be computed, for example using a formula derived from the product representation for the sine function:
\begin{equation}
\sin\theta = \theta \prod_{n=1}^\infty \left(1-\frac{z^2}{\pi^2n^2}\right) \Rightarrow \sum_{n=1}^\infty \frac1{\pi^2n^2+\theta^2} = \frac{\coth\theta}{2\theta} - \frac1{\theta^2} \;,
\end{equation}
where we have taken the logarithm of both sides of the first identity, and subsequently taken the derivative with respect to $\theta$. Applying this to our sum and performing the indefinite integral in $\mu$, we find
\begin{equation}
\sum_{n=-\infty}^{+\infty} \ln\left(4\pi^2T^2n^2+\vec k^2+m^2-\frac A{d-1}\right) = 2\ln\sinh\frac{\sqrt{\vec k^2+m^2-\frac A{d-1}}}{2T} + C \;,
\end{equation}
where $C$ is a constant of integration. By considering the $T\rightarrow0$ limit, one can show that it must be equal to $2\ln2$. This gives the result that
\begin{equation}
\tr\ln\left(-\partial^2+m^2-\frac A{d-1}\right) = 2T\int \frac{d^{d-1}k}{(2\pi)^{d-1}} \ln2\sinh\frac{\sqrt{\vec k^2+m^2-\frac A{d-1}}}{2T} \;.
\end{equation}
This can be split into the $T=0$ contribution and a finite temperature correction:
\begin{equation}
\tr\ln\left(-\partial^2+m^2-\frac A{d-1}\right) = \left.\tr\ln\left(-\partial^2+m^2-\frac A{d-1}\right)\right|_{T=0} + 2T\int \frac{d^3k}{(2\pi)^3} \ln\left(1-\exp-\frac{\sqrt{\vec k^2+m^2-\frac A3}}{2T}\right) \;,
\end{equation}
where we have set $d=4$ in the temperature correction.

The second expression can be computed in an analogous way, except that the numerator in the sum is a fourth-order polynomial, and it is thus necessary to split the fraction in partial fractions. We find the temperature correction to the trace to be equal to
\begin{equation}
2T\int \frac{d^3k}{(2\pi)^3} \left(\ln\left(1-\exp-\frac{\sqrt{\frac\alpha2+\frac{\sqrt{\alpha^2-4\beta}}2}}T\right) + \ln\left(1-\exp-\frac{\sqrt{\frac\alpha2-\frac{\sqrt{\alpha^2-4\beta}}2}}T\right)\right) \;,
\end{equation}
where we have used the short-hand notations $\alpha=m^2+2\vec k^2-A/3$ and $\beta=\vec k^2(\vec k^2+m^2+A)$.

\section{Minimizing the potential} \label{min}
From \cite{wij} we already know the zero-temperature effective potential. If we add the temperature correction found above, we can start the work of searching for minima. As the expressions involved are pretty much unhandleable, we use two strategies: expanding in series gives some analytical insight in the low- and high-$T$ behavior, and numerical minimization gives a global view of the temperature dependence. For the numerical part, we have used $\bar\mu^2 = \unit{4.12}{\lms^2}$, the value of $g\sigma'$ in the non-perturbative minimum at zero temperature\cite{Verschelde:2001ia}, and $N = 2$. It is possible to have $\bar\mu^2$ shift as $g\sigma'$ gets modified at finite temperature; this, however, does not significantly change the results.

\subsection{Numerical minimization}
Plotting and visually inspecting the potential reveals only one minimum, which coincides with the already known non-perturbative minimum at $T=0$. One would expect the zero-temperature perturbative solution to become a saddle-point of the potential at finite $T$, but it turns out that this saddle-point can only be found from $T=\unit{0.45}{\lms}$ onwards. For lower temperatures it seems that the saddle-point is located in a region of the parameter space where the effective potential has an imaginary part. For slightly higher temperatures, the saddle-point and the non-perturbative minimum merge and from a temperature of $\unit{0.67}{\lms}$ onwards no solutions to the gap equation can be found anymore. We will say more about this in paragraph \ref{hoget}.

\begin{figure}\begin{center}
\includegraphics[width=7cm]{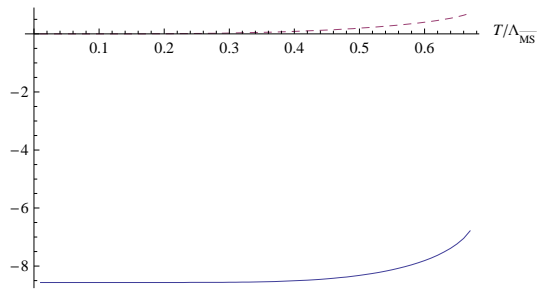}
\caption{The $\langle g^2A_\mu^2\rangle$ condensate (full line) and the asymmetry $\Delta_{A^2}$ (dashed line) as functions of the temperature, in units $\Lambda_{\overline{\text{MS}}}$. \label{eerstefiguur}}
\end{center}\end{figure}

The values of the condensates in the non-perturbative minimum are plotted in figure \ref{eerstefiguur}. In this figure we have used the sign and prefactor conventions of \cite{Chernodub:2008kf} instead of those from \cite{Verschelde:2001ia}, which means that the value of $\langle A_\mu^2\rangle$, being the opposite of $\sigma$, is negative. We see that the absolute value of $\langle A_\mu^2\rangle$ is slightly lowered by raising the temperature. The asymmetry is positive and rising, just as was found on the lattice in \cite{Chernodub:2008kf}. Our value for the asymmetry seems to be slightly lower, but as we have only done a one-loop calculation, one cannot expect the results to have very high accuracy.

\begin{figure}\begin{center}
\includegraphics[width=7cm]{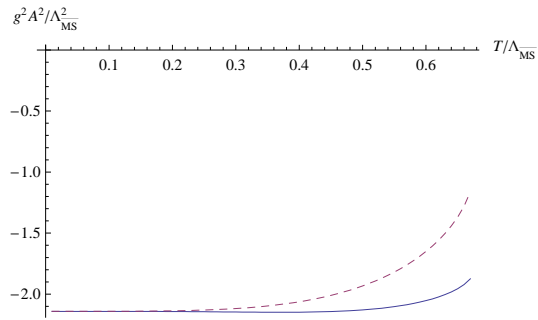}
\caption{The electric (dashed) and magnetic (full line) components of the $\langle A^2_\mu\rangle$ condensate as function of the temperature. (The magnetic component has been divided by three to be able to compare with the electric component.) \label{elem}}
\end{center}\end{figure}

In figure \ref{elem} the values of the electric part and the magnetic part are plotted separately. At $T=0$ both are, naturally, equal. When increasing the temperature, the electric component goes up, while the magnetic component remains approximately constant. This is also what has been found on the lattice\cite{maxim}. Similar conclusions for correlations in the gluon condensate were also found in \cite{ddm}.

\subsection{Low temperatures}
Analytically, the limit $T\sim0$ can be considered. In order to find the dominant behavior, we proceed as follows: First, the exponentials in the integrals are small for small $T$, meaning that $\ln(1-\exp x) \sim -\exp x$. Then, we expand the square roots for small $|\vec k|$. The expansion of the square root will only be valid up to a certain value of $|\vec k|$, but this is still a legitimate step as the greater values of $|\vec k|$ hardly contribute due to exponential suppression. In the expansion we only keep the terms up to the first non-trivial order of $|\vec k|$, after which the integrals can be easily evaluated.

We find that the three integrals in the potential have lowest-order behavior
\begin{equation}
-(N^2-1) \frac{m^{3/2}T^{5/2}}{2^{1/2}\pi^{3/2}}e^{-m/T} \;,\qquad -(N^2-1) \frac{\pi^2T^4}{90} \left(\frac{m^2-\frac A3}{m^2+A}\right)^{3/2} \;,\qquad -(N^2-1) \frac{T^{5/2}}{2^{3/2}\pi^{3/2}} \frac{(m^2+\frac A3)^{9/4}}{(m^2-\frac{5A}3)^{3/2}}e^{-\sqrt{m^2-\frac A3}/T} \;,
\end{equation}
respectively. It is clear that, for low $T$, the second integral will dominate. If we take this to be the first low-temperature correction, we find for the asymmetry
\begin{equation}
A = -\frac{g^2N\pi^2}{15m^2} \left(1-\frac{85}{1044}\frac{g^2N}{(4\pi)^2}\right) T^4 \;,\qquad \Delta_{A^2} = (N^2-1) \frac{g^2\pi^2}{30m^2} \left(1-\frac{85}{1044}\frac{g^2N}{(4\pi)^2}\right) T^4 \;,
\end{equation}
and (to this order in the temperature) there is no correction to $\langle g^2A_\mu^2\rangle$. If we apply a fit to the low-temperature part of our numerical data, the two results are in nice agreement. Mark that, as there is no $T^4$ correction to $\langle g^2A_\mu^2\rangle$ but there is a positive one to $\Delta_{A^2}$, the magnetic component of the condensate will decrease its value, or increase its absolute value, as can be seen in figure \ref{elem}. This is in opposition to the behavior of the electric component, which only decreases in absolute value. The increase in $|\langle g^2A_E^2\rangle|$ is small, however, and it is not sure how higher-loop correction will influence this result.

In \cite{Chernodub:2008kf} it was found that the value of the asymmetry was best described by an exponential form
\begin{equation}
\Delta_{A^2} \approx c T^2 e^{-m/T} \;,
\end{equation}
with $m=\unit{201(8)}{MeV}$. They, however, only had data for $T>0.4\;T_c$. For such higher temperatures, the lowest order in the expansion is, of course, not sufficient, and the exponential corrections cannot be ignored anymore. In order to investigate the behavior of the asymmetry, the numerical results have to be used again. It turns out to be very difficult to find a fit good enough in broader intervals. Given the complexity of the analytical expressions and given the fact that we have only done the calculations up to one-loop order, it is not possible to say more about it, however.

Mark that, given the fact that our model has a mass gap, one would expect an exponential behavior with the effective gluon mass in the exponential. Our not finding this is due to the Landau gauge prescription: the last inverse propagator in \eqref{tracespols} does not correspond to a simple Yukawa form when the asymmetry becomes nonzero. One should mark that the Landau gauge is singled out as the gauge where $\langle A_\mu^2\rangle$ reaches its minimum along the gauge orbit, giving this condensate a physical meaning. This is not the case with the asymmetry, and as such it is not all that surprising to find a non-exponential behavior for it.

\subsection{High temperatures} \label{hoget}
In order to get more insight in the disappearance of all solutions to the gap equation at higher $T$, we will expand the effective potential in this limit. Herefore it is necessary we return to \eqref{beginnetje} and do the integrations first, receiving, for example, for the first expression
\begin{equation}
-\frac{N^2-1}2 \frac{d-2}{(4\pi)^{(d-1)/2}} \Gamma(-\tfrac{d-1}2) T \sum_{n=-\infty}^{+\infty} \left(4\pi^2T^2n^2+m^2-\frac A{d-1}\right)^{(d-1)/2} \;.
\end{equation}
This can be expanded in high $T$, but one sees that the $n=0$ term has to be split off. We find:
\begin{eqnarray}
\sum_{n=-\infty}^{+\infty} \left(4\pi^2T^2n^2+m^2-\frac A{d-1}\right)^{(d-1)/2} &=& \left(m^2-\frac A{d-1}\right)^{(d-1)/2} + 2\sum_{n=1}^\infty \sum_{i=0}^\infty \begin{pmatrix} \frac{d-1}2 \\ i \end{pmatrix} \left(m^2-\frac A{d-1}\right)^i (4\pi^2T^2n^2)^{\frac{d-1}2-i} \nonumber \\
&=& \left(m^2-\frac A{d-1}\right)^{(d-1)/2} + 2(2\pi T)^{d-1}\sum_{i=0}^\infty \begin{pmatrix} \frac{d-1}2 \\ i \end{pmatrix} \left(\frac{m^2-\frac A{d-1}}{4\pi^2T^2}\right)^i \zeta(2i-d+1) \;,
\end{eqnarray}
where $\zeta(s) = \sum_{n=1}^\infty \frac1{n^s}$ is the Riemann zeta function. The term with $i=2$ will give a pole in the $d\rightarrow4$ limit. The second contribution from \eqref{beginnetje} can be expanded analogously.

All together, we find the following high-$T$ expansion:
\begin{multline} \label{hogeteenloop}
\frac{N^2-1}{2g^2N} \left(\frac9{13} m^4 + \frac13 A^2\right)
+ \frac{N^2-1}8 \left(m^2+\frac A3\right) T^2 - \frac{N^2-1}{12\pi} \left((m^2+A)^{\frac32} + \left(m^2-\frac A3\right)^{\frac32}\right) T \\ + \frac{N^2-1}{32\pi^2} \left(3m^4+\frac79A^2\right) \ln\frac{2\pi T}{\bar\mu} - \frac{N^2-1}{24\pi^2} \left(\frac{327}{208}m^4+\frac23m^2A-\frac{3623}{4176}A^2\right) \\
+ \frac{\zeta(3)(N^2-1)}{20736\pi^4} (81m^6-27m^4A+36m^2A^2+2A^3) T^{-2} + \cdots \;,
\end{multline}
where we have dropped the $T^4$ term, as it does not depend on the fields in any way and is, thus, irrelevant. Now one has to keep in mind that, at high temperatures, one expects the fields to scale with the temperature, and that all terms in the expansion above are effectively of the same order in $T$. However, one expects to have that $m^2\sim A\sim g^2T^2$, making the above series one in $g$, with the first two terms being of the same order. Solving the gap equation perturbatively yields at lowest order
\begin{equation}
m^2 = -\frac{13N}{72} g^2T^2 \ , \qquad A = -\frac N8 g^2T^2 \;,
\end{equation}
and for the condensates:
\begin{equation}
\langle g^2A_\mu^2\rangle = \frac{N^2-1}4 g^2T^2 \ , \qquad \Delta_{A^2} = \frac{N^2-1}{12} g^2T^2 \;.
\end{equation}
This is exactly the result one expects from a perturbative computation.

Going to higher order in this expansion, some subtleties are encountered. First note that the effective gluon masses squared are negative, and the next term in the expansion contains square roots of the masses. This gives an imaginary part to the potential. Another point of note is the fact that our expansion has effectively become a series in the coupling $g$ instead of one in the temperature. This means that, when going to higher order in $1/T$, one has to take into account the effect of higher-loop diagrams. These two problems are actually related.

\begin{figure}\begin{center}
\includegraphics[width=4cm]{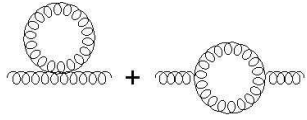}
\caption{The diagrams giving the Debye mass in hard-thermal-loop resummation. The ghost loop is not necessary\cite{braatenpisarski}. \label{debye1}}
\end{center}\end{figure}

\begin{figure}\begin{center}
\includegraphics[width=4cm]{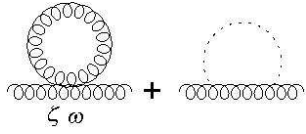}
\caption{More diagrams that need to be resummed: two coming from the $\sigma$ part of the LCO Lagrangian, and again the same two diagrams from the $\varphi_{\mu\nu}$ part of the LCO Lagrangian. The index ``$\zeta$, $\omega$'' has been added to remind of the fact that this depicts the two four-gluon vertices coming from the LCO formalism. The dotted line is the $\sigma$ or $\phi_{\mu\nu}$ propagator. \label{debye2}}
\end{center}\end{figure}

It has been known for a long time that, at higher temperatures, the perturbation series must be reorganized.\footnote{See for example \cite{andersenstrickland}.} In ordinary pure Yang--Mills theory, this amounts to giving the timelike gluon a Debye mass $m_D^2 = \tfrac N3g^2T^2$, which effectively resums the hard (high momentum) contributions of the diagrams in figure \ref{debye1}. In our formalism, however, there are four additional vertices. This gives rise to four extra diagrams that need to be resummed. They are shown in figure \ref{debye2}. Computing these diagrams, it turns out that they exactly cancel the lowest-order contribution from the condensate.

When doing this resummation, one has to watch out for double counting, which can happen when considering diagrams without external lines\cite{braatenpisarski}. However, it turns out that this double counting is put right by the mismatch in symmetry factors in the diagrammatic expansion of the vacuum energy. As such, we can proceed without worrying about this. Adding the resummed diagrams to the result found in \eqref{hogeteenloop}, we find up to the effective order $g^3$:
\begin{multline}
\frac{N^2-1}{2g^2N} \left(\frac9{13} m^4 + \frac13 A^2\right)
+ \frac{N^2-1}8 \left(m^2+\frac A3\right) T^2 \\ - \frac{N^2-1}{12\pi} \left(\left(m^2+\frac{13N}{72} g^2T^2+A+\frac N8 g^2T^2+m_D^2\right)^{\frac32} + \left(m^2+\frac{13N}{72} g^2T^2-\frac A3-\frac N{24}g^2T^2\right)^{\frac32}\right) T + \cdots \;,
\end{multline}
where, again, terms not containing the fields $m^2$ and $A$ have been dropped. Again solving this perturbatively, we find:
\begin{equation}
m^2 = -\frac{13}{18} g^2N\left(\frac{T^2}4 - \frac{m_DT}{4\pi} + \cdots\right) \ , \qquad A = -\frac32 g^2N \left(\frac{T^2}{12} - \frac{m_DT}{36\pi} + \cdots \right) \;,
\end{equation}
and analogously for $\langle g^2A_\mu^2\rangle$ and $\Delta_{A^2}$. This is exactly what one would expect from perturbation theory.

\section{Conclusions} \label{conc}
We computed the effective action of SU($N$) Landau gauge Yang-Mills theory in the presence of a dimension two condensate and an asymmetry in this condensate. Figure \ref{eerstefiguur} is the main result of this article. We find good qualitative agreement with the numerical results of \cite{Chernodub:2008kf}, with some discrepancies due to different definitions --- we define the condensates with the perturbative contributions subtracted, whence they vanish in the high temperature, perturbative, regime. The quantitative agreement is less excellent, but as our computations are just one-loop and the coupling is not all that small, one may not hope for miracles. A two-loop treatment, however, is intractable, even at zero temperature\cite{john}.

The low-$T$ behavior seems to be best described by $\Delta_{A^2} = \alpha T^4$, as a naive computation in an Abelian Higgs model would lead us to expect\cite{Chernodub:2008kf}. The mismatch with the exponential fit found in \cite{Chernodub:2008kf} is probably due to their having data only for $T>0.4\;T_c$. At high temperatures it turns out that resummation \emph{\`a la} hard-thermal-loop is necessary, and not doing this will give no solutions when imposing that the effective action be real. After resumming the necessary diagrams, the perturbative values for the condensates are recovered. No non-perturbative solutions are found. The only part of the temperature range where we cannot boast good results is around the phase transition. At that point the temperature is already too high to trust a simple one-loop computation, and the high-temperature expansion cannot be expected to still yield good results at a temperature so low.

When this article was in preparation, we learned that lattice computations for the full dimension two condensate and for the electric and magnetic components separately at finite temperature have been completed\cite{maxim}. Qualitative agreement is again good. For $T<T_c$, it is indeed found that the electric component shows much more temperature dependence than the magnetic component, which is nearly constant in that range.

\section*{Acknowledgements}
We wish to thank M.~Chernodub for encouraging this research and for many discussions. D.V. also gratefully acknowledges helpful discussions with D.~Dudal.

\end{document}